# Non-equilibrium Phonon Thermal Resistance at MoS$_2$/Oxide and Graphene/Oxide Interfaces

Weidong Zheng[1,2], Connor J. McClellan[3], Eric Pop[3,4], Yee Kan Koh[1,2*]

[1]*Department of Mechanical Engineering, National University of Singapore, Singapore 117576*

[2]*Centre for Advanced 2D Materials, National University of Singapore, Singapore 117542*

[3]*Department of Electrical Engineering, Stanford University, Stanford, California 94305, USA*

[4]*Department of Materials Science and Engineering, Stanford University, Stanford, California 94305, USA*

*Corresponding Author Email: mpekyk@nus.edu.sg

# ABSTRACT

Accurate measurements and physical understanding of thermal boundary resistance ($R$) of two-dimensional (2D) materials are imperative for effective thermal management of 2D electronics and photonics. In previous studies, heat dissipation from 2D material devices was presumed to be dominated by phonon transport across the interfaces. In this study, we find that in addition to phonon transport, thermal resistance between non-equilibrium phonons in the 2D materials could play a critical role too when the 2D material devices are internally self-heated, either optically or electrically. We accurately measure $R$ of oxide/MoS$_2$/oxide and oxide/graphene/oxide interfaces for three oxides (SiO$_2$, HfO$_2$, Al$_2$O$_3$) by differential time-domain thermoreflectance (TDTR). Our measurements of $R$ across these interfaces with external heating are 2-to-4 times lower than previously reported $R$ of the similar interfaces measured by Raman thermometry with internal self-heating. Using a simple model, we show that the observed discrepancy can be explained by an additional internal thermal resistance ($R_{int}$) between non-equilibrium phonons present during Raman measurements. We subsequently estimate that for MoS$_2$ and graphene, $R_{int} \approx 31$ and $22$ m$^2$ K GW$^{-1}$, respectively. The values are comparable to the thermal resistance due to finite phonon transmission across interfaces of 2D materials and thus cannot be ignored in the design of 2D material devices. Moreover, the non-equilibrium phonons also lead to a different temperature dependence than that by phonon transport. As such, our work provides important insights into physical understanding of heat dissipation in 2D material devices.

# INTRODUCTION

Two-dimensional (2D) materials, e.g., graphene and transition metal dichalcogenides (TMDs), are widely investigated for applications in nanophotonics[1] and nanoelectronics.[2] In such applications, thermal management of 2D material devices is crucial to avoid performance degradation due to over-heating.[3] Most 2D materials have highly anisotropic thermal conductivity, with a lower thermal resistance along the basal plane.[4-6] As a result of the high in-plane thermal conductivity, heat dissipation from 2D material devices with sufficiently small lateral dimensions is governed mainly by heat transport across their 2D interfaces with adjacent insulators or metal contacts,[7-9] through an interfacial property called the thermal boundary resistance ($R$). (We note that for interfaces of 2D materials of few atomic layers, most researchers in thermal[10-15] and device[16-18] community used the terms "thermal boundary resistance" or "interfacial thermal resistance" to refer to the total thermal resistance between the heat source in the 2D material and the adjacent materials, not just the phononic thermal resistance across the interfaces. We follow this definition in this paper.) Hence, accurate measurements of the thermal boundary resistance of 2D materials are essential for effective thermal management of 2D material devices.

Over the past decade, the thermal boundary resistance of 2D materials has been examined through measurements by time-domain thermoreflectance (TDTR),[19-24] Raman thermometry,[13-18] pump-probe transient absorption[10] and the 3ω method.[25] Surprisingly, there is a large discrepancy in the literature on the thermal boundary resistance of similar interfaces of 2D materials, especially when $R$ is measured by different techniques. For example, Koh *et al.*[23] and Chen *et al.*[25] found that $R$ of graphene/SiO$_2$ interfaces, measured by TDTR and the 3ω method, is only ≈ 11 m$^2$ K GW$^{-1}$, ≈4 times smaller than that measured by Vaziri *et al.*[18] and Freitag *et al.*[16] using the electrical heating Raman thermometry. Interestingly, despite the discrepancy, finite elastic and inelastic phonon transmission[14, 17-19] across the 2D material interfaces is presumed to be solely responsible for the observed thermal boundary resistance in all prior studies. Often, the disparity in the thermal resistance values is attributed to different transmission probability of phonons at the interfaces due to the presence of capping layer[26, 27] and/or different interface quality,[17, 24] as a small change in the van der Waals gap between 2D materials and substrates could affect the interfacial thermal transport.[28]

In addition to the aforementioned explanations, another possible explanation to the discrepancy in the measured $R$, which was ignored in most prior analysis, is that phonons are not in local thermal equilibrium during Raman measurements. In typical Raman measurements, 2D materials are heated by laser irradiation[13-15] or electrical heating[16-18] while the steady-state temperatures of vibrational modes in the 2D materials are monitored from either shifts of the Raman peaks or intensity ratios of Stokes and anti-Stokes peaks. Thus, an implicit assumption of the Raman measurements is that different phonons in the 2D materials are in local thermal equilibrium. When the phonons monitored by Raman are out of thermal equilibrium with the phonons responsible for heat transfer across the 2D interface, an additional internal thermal resistance ($R_{int}$) between these phonon populations could lead to a larger effective thermal boundary resistance in Raman measurements.[11-13] This is similar to how non-equilibrium phonons in prior Raman measurements of the *in-plane* thermal conductivity of 2D materials have led to underestimation of the derived *in-plane* thermal conductivity, as discussed extensively by Vallabhaneni *et al.*[29], Lu *et al.*[30] and Sullivan *et al.*[31] We note that in 2D material devices, heat is usually generated through thermalization and scattering of hot electrons within the 2D materials upon optical or electrical excitation,[32] similar to heating of 2D materials in Raman measurements. Thus, heat dissipation in the 2D material devices could be affected by $R_{int}$ as well.

Despite prior discussion on the effects of $R_{int}$,[11-13] the role of $R_{int}$ in thermal boundary resistance of 2D materials has yet to be experimentally demonstrated up to now. For example, in Ref. 13, researchers found that the measured $R$ of $MoS_2/SiO_2$ interfaces is much higher than that of the $Au/Ti/MoS_2/SiO_2$ interfaces reported in Ref. 33. While $R_{int}$ could be a valid explanation, the difference in $R$ could also originate from the fact that Ti reacts with $MoS_2$[34, 35] and thus the $R$ of the $Ti/MoS_2$ interfaces might not be intrinsic. Hence, one key challenge for experimental demonstration of the role of $R_{int}$ is that the interfaces of the 2D materials measured by different techniques must have similar qualities, because the thermal boundary resistance is strongly affected by the quality of the interfaces, e.g., the conformity of the 2D materials to the substrate (i.e., the amount of voids at the interfaces),[24] the amount of residues,[36] and how the 2D materials were transferred/synthesized.[17]

In this study, we provide experimental evidence to suggest that internal thermal resistance plays a critical role in the effective thermal resistance of Raman measurements and hence heat transport across interfaces of 2D materials. We accurately measure $R$ of oxide/$MoS_2$/oxide and oxide/graphene/oxide interfaces for some common oxides (i.e., $SiO_2$, $HfO_2$, $Al_2O_3$), by differential time-domain thermoreflectance (TDTR) which we have developed. For fair comparison with prior Raman measurements, we ensure that our $MoS_2$ samples were prepared by the same research group using similar procedures, to ensure similar interface qualities for our samples. Interestingly, even for samples with similar quality, we observe that our measurements of $R$ across these interfaces of 2D materials are substantially lower than previously reported $R$ of similar interfaces measured by Raman thermometry. Through a simple model that takes into consideration $R_{int}$, we find that the discrepancy could be explained by the presence of non-equilibrium phonon thermal resistance at 2D material interfaces. We further estimate that $R_{int} \approx$ 31 and 22 m$^2$ K GW$^{-1}$ for $MoS_2$ and graphene, respectively. In addition, the non-equilibrium phonons are also responsible for the difference in the temperature dependence of $R$ between prior Raman measurements and that due to elastic phonon transmission.

## RESULTS AND DISCUSSION

Our samples consist of transferred single-layer $MoS_2$ and graphene (Gr) grown by chemical vapor deposition (CVD) and sandwiched between two layers of ≈14-nm-thick oxide films on GaN/sapphire substrate, see Fig. 1(a). We carefully select three oxides (i.e., $SiO_2$, $HfO_2$ and $Al_2O_3$) that do not react with $MoS_2$ or graphene to ensure the intrinsic interfaces of 2D materials. The bottom and top $SiO_2$ films were deposited by radio frequency (rf) magnetron sputtering and electron beam evaporation, respectively, while the $HfO_2$ and $Al_2O_3$ films were deposited by atomic layer deposition (ALD). Details of the sample preparation are summarized in the Methods section. To check the quality of our transferred $MoS_2$ and graphene, we first measured the Raman spectra of $MoS_2$ and graphene transferred on the oxide substrates excited with a 532 nm wavelength continuous laser, before and after (as labelled in Fig. 1(b) and 1(c)) encapsulation of the top oxide layer. For oxide/$MoS_2$/oxide samples, although we find $MoS_2$ $E'$ peak broadening after the top oxide encapsulation, we do not observe the appearance of LA(M) peak, suggesting that the disorder within $MoS_2$ induced by encapsulation with the oxide layer is limited and not that defective (interdefect distance > 3.2 nm).[37, 38] We also observe a red shift and peak broadening for the $A_1'$ mode, which could be due to the doping effect.[39] For the oxide/Gr/oxide samples, we observe no significant D peaks, indicating that any damage induced by deposition of the oxide films is below the Raman detection limit.[8, 24] We also notice that after growing the encapsulating oxide film, the graphene samples show a red-shift of the G peak position; the red-shift could be due to doping of graphene during the oxide deposition process.[40, 41] We also employed tapping-mode atomic force microscopy (AFM) to characterize the cleanness and conformity of $MoS_2$ and graphene after the transfer process. Little residues are observed in the topographic images of our $MoS_2$ and graphene on oxide films in Fig. S1.

Using an approach similar to that described in Ref. 24, we find that all samples conform to the oxide films with a percentage of contact area of ≈100%, see Fig. 1(d).

We measure the thermal resistance of oxide/MoS$_2$/oxide ($R_{OMO}$, where O and M denote oxide and MoS$_2$, respectively) and oxide/Gr/oxide ($R_{OGO}$, where G denotes graphene) double interfaces by differential TDTR. Details of our implementation are discussed in the Methods section and Supporting Information section S2. We employ differential TDTR instead of conventional TDTR to achieve better accuracy; the uncertainty of our measurements is ≈8%, much smaller than that of prior measurements.[13, 17, 23, 25] For better comparison with Raman measurements, we derive $R$ of single MoS$_2$/oxide and Gr/oxide interfaces, $R_{MO}$ and $R_{GO}$, from our measured $R_{OMO}$ and $R_{OGO}$ of the double interfaces, by assuming that the thermal resistances of the top and bottom MoS$_2$/oxide (or Gr/oxide) interfaces add in series,[8] i.e., $R_{MO} = 0.5 \times R_{OMO}$ and $R_{GO} = 0.5 \times R_{OGO}$. For the rest of the paper, we focus on discussing the thermal resistance of the single interfaces, not of the double interfaces.

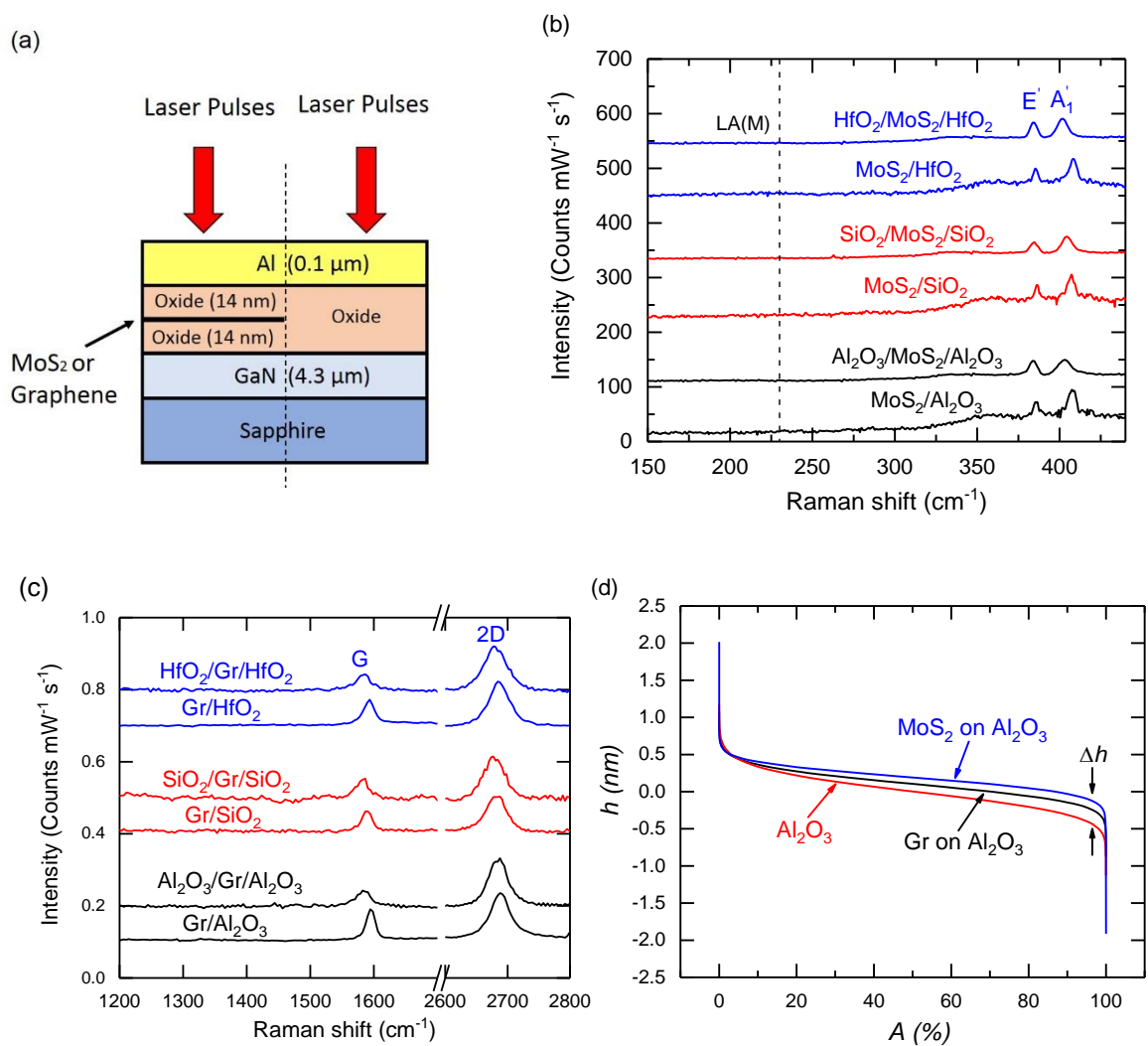

**Fig. 1. (a)** Cross-sectional schematic of our oxide/MoS$_2$/oxide and oxide/Gr/oxide samples. For differential TDTR measurements, some control areas were intentionally left without MoS$_2$ or graphene during the preparation of our samples. Raman spectra of **(b)** MoS$_2$ and **(c)** graphene transferred on oxides before (labelled as "MoS$_2$/oxide" or "Gr/oxide") and after (labelled as "oxide/MoS$_2$/oxide" or "oxide/Gr/oxide") encapsulation of a ≈14-nm-thick oxide film. All spectra are shifted vertically for ease of comparison. **(d)** Determination of the conformity of MoS$_2$ and graphene on Al$_2$O$_3$ film from plots of relative height $h$ (derived from Supporting Information Fig. S1) as a function of accumulative percentage area $A$. MoS$_2$ (or graphene) is assumed to be conformal when the difference in the relative height of MoS$_2$ (or graphene) and oxide substrate $\Delta h < 0.7$ nm (or 0.5 nm for graphene).

In Fig. 2, we plot our measured $R$ of Gr/oxide and MoS$_2$/oxide interfaces as a function of effective Debye temperature ($\Theta_{\text{eff}}$) of substrates, and compare the measurements to the estimation from the phonon radiation limit[42] ($R_{\text{rad}}$) of the corresponding interfaces. (See Section S6 in Supporting Information for details of calculations of $R_{\text{rad}}$.) We define the effective Debye temperature as[43] $\Theta_{\text{eff}} = \frac{\hbar v}{k_B}(6\pi^2 \frac{N}{V})^{\frac{1}{3}}$, where $N/V$ is the atom density, $v = \left[\frac{1}{3}\left(\frac{2}{v_T^3} + \frac{1}{v_L^3}\right)\right]^{-1/3}$ is the average sound velocity, $v_L$ and $v_T$ are the sound velocity of longitudinal (LA) and transverse acoustic (TA) phonons in the oxides or metals, respectively, $\hbar$ is the reduced Planck constant and $k_B$ is the Boltzmann constant. Different from the common practice, we assume an atom density of $N/V = 7.0\times10^{28}$ m$^{-3}$ (estimated from the average atom density of all substrates in Fig. 2) and $v_T = 0.5\times v_L$ for all substances; in this way, $\Theta_{\text{eff}}$ only depends on $v_L$, and thus calculations of $R_{\text{rad}}$ can be plotted as continuous lines in Fig. 2. We fit our measured $R$ with $3R_{\text{rad}}$ for MoS$_2$ interfaces and $2R_{\text{rad}}$ for graphene interfaces. Similar to our previous observation,[19] we observe that $R$ of the interfaces is larger than that determined from the phonon radiation limit, suggesting that heat transport across MoS$_2$/oxide and Gr/oxide interfaces is mainly due to the elastic transmission of phonons.[19]

We further compare our measured $R$ of MoS$_2$/oxide and Gr/oxide interfaces to prior measurements by Raman thermometry (blue),[13-18] pump-probe transient absorption (magenta),[10] the 3$\omega$ method (black),[25] and TDTR (red).[19-23] For Gr/SiO$_2$ interfaces, our measured $R$ is consistent with previous measurements by the 3$\omega$ method[25] and TDTR[34], but substantially lower than prior measurements by Raman thermometry,[16, 18] see Figs. 2(a) and 3(a). For MoS$_2$ interfaces, however, we observe a wide spread of measured $R$ values even when the samples were prepared using similar procedures and measured by similar techniques, see Fig. 2(b). We note that a larger-than-intrinsic $R$ value could be measured when there are sample preparation and measurement errors, such as a low degree of conformity for the MoS$_2$ layer,[24] the existence of residues after layer transfer or exfoliation, improper subtraction of the substrate thermal resistance, and incorrect assumptions of the laser percentage absorbed in Raman measurements with optical heating.[13]

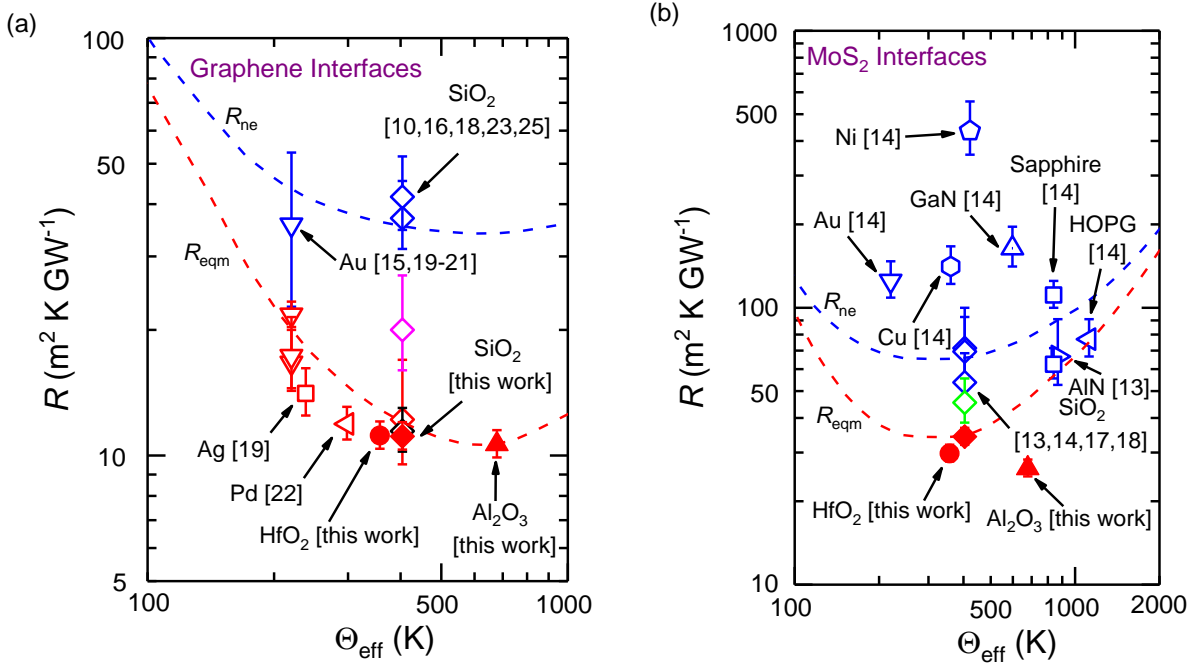

**Fig. 2.** Comparison of thermal resistance $R$ of interfaces of **(a)** graphene and **(b)** MoS$_2$, with metals or oxides as labelled, measured in this work (solid symbols) and prior studies (open symbols), as a function of effective Debye temperature $\Theta_{eff}$ of metal or oxide. The black, red, magenta and blue symbols are measurements by the 3ω method, TDTR, pump-probe transient absorption and Raman thermometry, respectively. The green symbol is a recent Raman measurement, in which MoS$_2$ layer was indirectly heated by Joule heating.[18] The red dashed lines are calculations of thermal resistance for graphene and MoS$_2$ in (a) and (b), respectively, when local equilibrium is established, while the blue dashed lines are calculations using Eq. (1).

Thus, throughout this paper, we compare our measurements and calculations primarily to the lower bound of the Raman measurements. Even when we only consider the lower bound values, the compiled data of Raman measurements are still 2-to-4 times larger than the TDTR or 3ω measurements. This large discrepancy between the TDTR (or 3ω) and the Raman measurements cannot be explained by measurement uncertainties. The only exception, represented by an open green diamond in Fig. 2(b), is a recent Raman measurement of heat transport across a heterostructure of Gr/MoS$_2$/SiO$_2$, in which the MoS$_2$ layer was indirectly heated by Joule heating in a graphene heater rather than directly heated by laser.[18]

Measurements by different techniques also exhibit different temperature dependence of $R$. In Figs. 3(a) and 3(b), we plot the $R$ of graphene/oxide and MoS$_2$/oxide interfaces measured by different techniques as a function of temperature ($T$). We find that, different from prior Raman measurements[13] but similar to data by the 3$\omega$ method,[25] our measured $R$ displays a weaker temperature dependence, with $R$ of the MoS$_2$/oxide interface plateauing when $T > 300$ K. The weak temperature dependence is consistent with the scenario in which heat is carried across the interfaces predominantly by elastic transmission of phonons.[44] In such a case, the temperature dependence of $R$ largely follows the temperature dependence of heat capacity of the side with lower Debye temperature (i.e., MoS$_2$ for the MoS$_2$/oxide interface),[44, 45] see Section S5 in Supporting Information for more details. Because the Debye temperature for acoustic phonons in MoS$_2$ is only $\approx$260 K,[46, 47] a strong temperature dependence is not expected for the MoS$_2$/oxide interface when $T > 300$ K, if heat transport across the interfaces is limited by elastic phonon transmission.

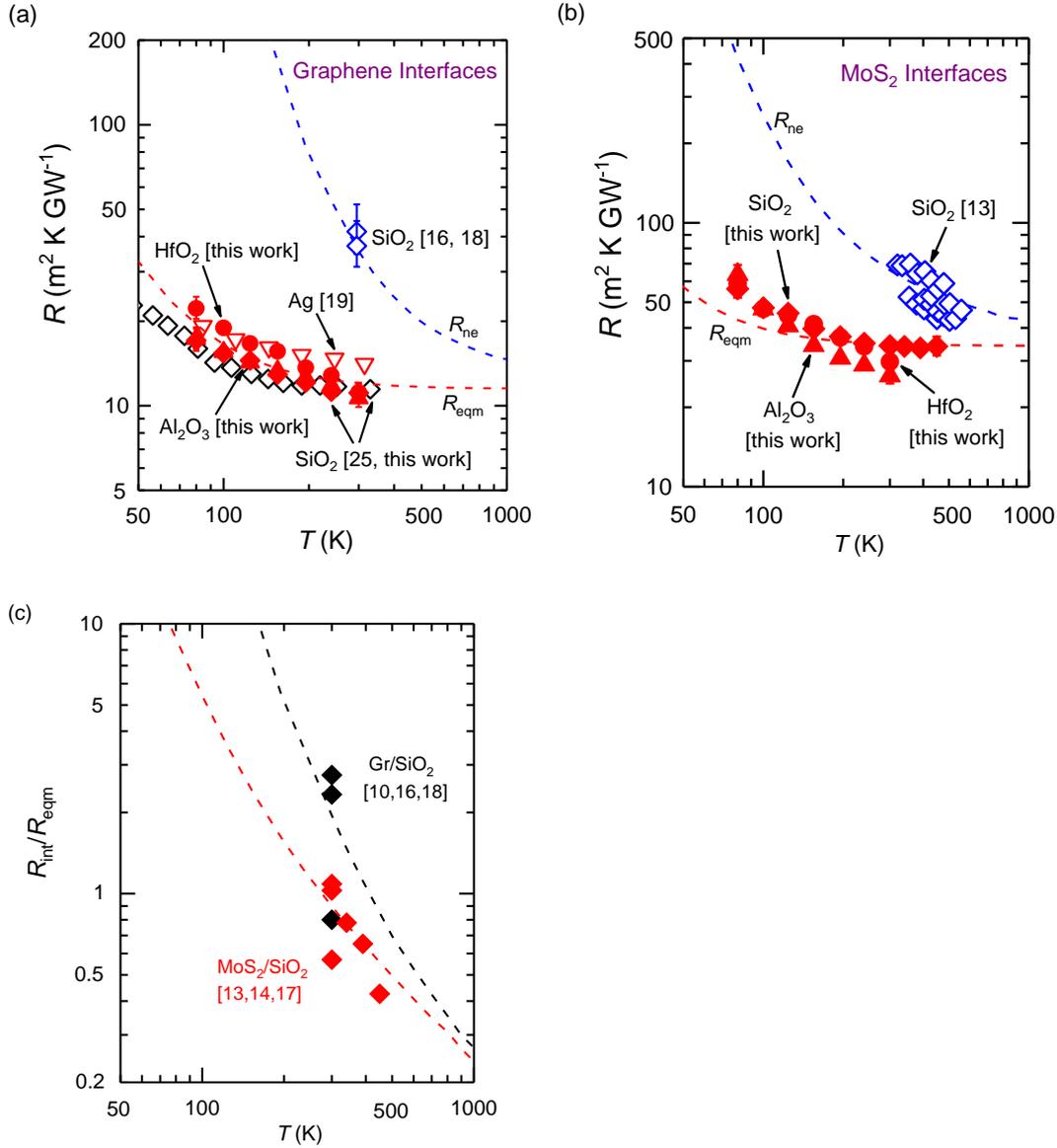

**Fig. 3.** Temperature dependence of $R$ of interfaces of **(a)** graphene and **(b)** MoS$_2$ on oxide or metal (as labelled), measured in this work (solid symbols) and prior studies (open symbols). The black, red, and blue symbols represent the measurements by the 3ω method, TDTR and Raman thermometry, respectively. The red dashed lines in (a) and (b) are calculations of $2R_{rad}$ for interfaces of Gr/SiO$_2$ and $3R_{rad}$ for interfaces of MoS$_2$/SiO$_2$, respectively. The blue dashed lines are the calculated $R_{ne}$ of Gr/SiO$_2$ and MoS$_2$/SiO$_2$ interfaces by adding the internal resistance between non-equilibrium phonons into the calculated $R_{eqm}$ for MoS$_2$ and graphene interfaces. (c) Ratios of $R_{int}$ to $R_{eqm}$ as a function of temperature for MoS$_2$/SiO$_2$ (red symbols) and graphene/SiO$_2$ (black symbols) interfaces. We estimate $R_{int}$ from the difference between our TDTR measurements (as $R_{eqm}$) and the Raman measurements (as $R_{ne}$, from Ref. 13, 14, 17 for MoS$_2$ and Refs. 10, 16, 18 for graphene, respectively). (For the Raman measurements from Ref. 13, we use the average value of $R_{ne}$ within the temperature range of ±5% from our TDTR

measurements.) The red and black dashed lines are the Eq. (1) calculations of $R_{int}/R_{eqm}$ for MoS$_2$ and graphene interfaces, respectively.

One possible explanation for the discrepancies we observe between our measurements and prior Raman measurements is that, unlike some previous measurements, we encapsulated the 2D materials with ≈14-nm-thick top oxide films. The encapsulating oxide layers could modify the phonon dispersion of the 2D materials, resulting in hybridization of phonons and emergence of Rayleigh wave modes that could facilitate heat transport across the interfaces of 2D materials.[33, 48] The top oxide layers might also induce repopulation of the flexural ZA phonons from other phonons[27] and thus enhance the interfacial heat transport, because heat is carried across the interfaces of 2D materials mainly by flexural ZA phonons.[26] Also, a limited amount of damage could be induced by the deposition of capping layer even though no LA(M) peak and D peak are observed in the Raman spectra of MoS$_2$ and graphene samples, and the defects at the interfaces could enhance interfacial thermal transport.[22] The explanation is, however, contradictory to recent measurements of $R$ of MoS$_2$/SiO$_2$ interfaces[13] and Gr/SiO$_2$ interfaces[16, 18], with and without a capping layer. In those studies, encapsulation had a negligible effect on the measured $R$ of both interfaces. Also, encapsulation should not affect the temperature dependence of $R$.

Another possible explanation to the observed discrepancies is that unlike TDTR or the 3$\omega$ measurements, different phonons in 2D materials are not in local thermal equilibrium during the Raman measurements of $R$ where 2D materials are directly heated by optical or electrical excitation.[11-13] In other words, the 2D materials are externally heated in TDTR measurements while internally heated in typical Raman thermometry measurements. Similar observations have been reported in the Raman measurements of *in-plane* thermal conductivity of suspended graphene.[29-31] In TDTR or measurements with an electric heater on top, energy from laser radiation or electric current is first transferred to electrons and subsequently to phonons in the conductive heater film. Usually, local thermal equilibrium is reached within this conductive films fairly easily due to strong electron-phonon coupling.[49] As a result, as long as a sufficiently low modulation frequency[50] and a sufficiently large heater size[51] are employed in TDTR measurements, heat transport across all the subsequent layers and interfaces is essentially diffusive with all phonons in local thermal equilibrium.[52]

On the other hand, in Raman measurements[13-15] with optical heating, the samples (e.g., 2D materials) are directly heated by a high-fluence laser beam, driving the excited electrons and phonons into non-equilibrium states. Moreover, temperature is determined through shifts of Raman peaks of optical phonons, which depend on the temperatures of phonons (e.g., LA and TA phonons) interacting with these Raman-active modes. (Raman shifts are mainly induced by enhanced anharmonic interactions of the monitored optical phonons due to increased populations of LA and TA modes involved in the anharmonic scattering. Readers are referred to Ref. 31, 53, 54 for more detailed discussions on what Raman shifts measure.) Thus, when phonons are not in local thermal equilibrium, the temperature that the Raman shifts measure could be substantially different from the temperature of the dominant heat carriers in the samples, resulting in an additional thermal resistance.

Let us consider Raman measurements of Gr/substrate interfacial thermal resistance, for example. In most prior Raman measurements,[15, 16] the temperature of the graphene was measured through shifts of the G peak of the graphene Raman spectrum. Because the G-mode LO phonons near the Brillouin zone center are mainly scattered by LA and TA phonons,[55, 56] the shifts of the G peak mainly reflect the temperatures of LA and TA phonons. However, heat is predominantly carried across graphene interfaces by flexural ZA phonons, not LA and TA phonons.[26, 57, 58] Thus, when ZA phonons are not in equilibrium with LA and TA phonons, an additional internal thermal resistance can develop between the ZA phonons and LA, TA phonons in Raman measurements.[11, 12] The local non-equilibrium of phonon modes occurs in 2D materials regardless of the laser power applied in Raman measurements, because the weak coupling between the LA/TA modes and the ZA modes persists regardless of the laser power. Thus, even when a low laser power is applied, the local non-equilibrium does not vanish. In fact, according to calculations by multi-temperature models, the local non-equilibrium is more pronounced at lower laser power because coupling between the LA/TA modes and the ZA modes is weaker at lower temperature,[31] and persists even when the laser power is as low as 10 μW.[29]

To test whether the discrepancies we observed between TDTR (or 3ω) and Raman measurements could be explained by the non-equilibrium phonons in Raman measurements, we consider the effect of local thermal non-equilibrium on the measured $R$, by deriving an equation relating (1) $R_{eqm}$, the thermal resistance of Gr/oxide or MoS$_2$/oxide interfaces measured by techniques where local equilibrium is established (e.g., TDTR and the 3ω method), (2) $R_{ne}$, the thermal resistance of the single interfaces measured by techniques in which the excited phonons are in strong non-equilibrium (e.g., Raman thermometry with electrical or optical self-heating and pump-probe transient absorption method), and (3) the internal thermal resistance ($R_{int}$) between the non-equilibrium phonons in the 2D materials. Note that $R_{eqm}$ corresponds to thermal resistance due to finite phonon transmission at the interfaces of 2D materials. Here, we assume that phonons are in non-equilibrium after laser irradiation or Joule self-heating in Raman measurements, and that heat flows to ZA phonons only through anharmonic scattering with in-plane acoustic modes (LA and TA phonons).[59] Thus, an internal resistance exists between LA and TA phonons, temperature of which the Raman experiments monitor, and ZA phonons that carry heat across the interfaces. We further assume that this internal thermal resistance and the thermal boundary resistance of 2D materials add in series, i.e.,

$$R_{ne} = R_{eqm} + R_{int} \qquad (1)$$

Here, $R_{int}$ is estimated[60, 61] from $R_{int} = \frac{\tau_{eff}}{C_{LT}}$, where $C_{LT}$ is the heat capacity per unit area of LA and TA phonons in single-layer graphene or MoS$_2$, and $\tau_{eff}$ is the effective relaxation time for the inelastic scattering process between in-plane acoustic modes and ZA phonons. We determine the value of $C_{LT}$ for single-layer graphene from calculations of a lattice dynamics model reported in Ref. 62, while for single-layer MoS$_2$, we calculate $C_{LT}$ using a 2D Debye model,[63] see S3 in Supporting Information for details of the 2D Debye model. We derive $\tau_{eff}$ at room temperature from the frequency-dependent relaxation times of MoS$_2$ and graphene, see Section S4 in Supporting Information for more details. For simplicity, we use $R_{eqm} = 2R_{rad}$ and $3R_{rad}$ in the calculations for graphene (Figs. 2(a) and 3(a)) and MoS$_2$ (Figs. 2(b) and 3(b)) interfaces, respectively; the expressions for $R_{eqm}$ are derived by fitting the TDTR measurements, see Fig. 2. We note that this simple model is only meant as an "order-of-magnitude" approximation of the values of $R_{int}$, and the accuracy of the model is not pivotal to the conclusions of this paper.

We compare calculations of $R_{ne}$ using Eq. (1) (blue dashed lines) to $R$ of MoS$_2$/oxide and Gr/oxide interfaces in Fig. 2. We find that the derived $R_{ne}$ agrees reasonably well with the lower bounds of the Raman measurements of both MoS$_2$ and graphene interfaces for a wide range of $\Theta_{eff}$, see Fig. 2, suggesting that the observed differences between TDTR (or 3ω) and Raman thermometry is consistent with the non-equilibrium phonon thermal resistance in 2D materials.

To understand the strong temperature dependence of the Raman measurements of $R$ of MoS$_2$ interfaces by Yalon *et al.*[13] in Fig. 3(b), we fit our calculations to the measurements and find that $R_{ne}$ could fit the measurements well when we assume $\tau_{eff}^{-1} \propto T$,[64, 65] see the blue dashed lines in Fig. 3 and Section S4 in Supporting Information for more details. This temperature dependence is typically found when phonon scattering is limited by the three-phonon Umklapp process. Because a similar temperature dependence is also found in other 2D materials, their interfaces could also exhibit a strong temperature dependence for $R$ whenever non-equilibrium phonon thermal resistance plays an important role.

Finally, we estimate the internal thermal resistance ($R_{int}$) due to non-equilibrium phonons from the differences between our TDTR measurements and the Raman measurements by Refs. 13, 14, 17 for $MoS_2$ and by Refs. 10, 16, 18 for graphene. At 300 K, $R_{int} \approx 31$ and 22 $m^2$ K $GW^{-1}$ for $MoS_2$ and graphene, respectively. The values of $R_{int}$ are comparable to our TDTR measurements, $R_{eqm}$, which accounts for the thermal resistance due to finite phonon transmission at the interfaces. We find that, at 300 K, $R_{int}/R_{eqm}$ ratios are 0.8 and 2.0 for the $MoS_2/SiO_2$ and $Gr/SiO_2$ interfaces respectively, see Fig. 3(c). The large $R_{int}/R_{eqm}$ ratios indicate that, whenever phonons are not in local thermal equilibrium, heat transport across interfaces of 2D materials is not solely controlled by interfacial phonon transmission, as commonly assumed in previous analysis.[14, 16, 17] Importantly, while $R_{int}/R_{eqm}$ ratios decrease with temperature, the ratios are still substantial at 450 K, see Figure 3(c). Thus, even at elevated temperatures, non-equilibrium phonon thermal resistance cannot be ignored in designs and analyses of thermal management of 2D material devices, because optically and electrically excited phonons in 2D material devices are usually out-of-equilibrium[29-31] with phonons that carry heat across their interfaces.

## CONCLUSIONS

In conclusion, we report the thermal resistance of $MoS_2$/oxide ($SiO_2$, $HfO_2$, $Al_2O_3$) and graphene/oxide interfaces over a temperature range of 80-450 K. The observed discrepancies between thermal resistance measured by TDTR (or $3\omega$) and Raman thermometry can be explained by a simple model that includes an additional internal thermal resistance from the non-equilibrium phonons. This non-equilibrium phonon thermal resistance also leads to a stronger-than-expected temperature dependence of thermal resistance in previous Raman measurements. Moreover, we find that the magnitude of non-equilibrium phonon thermal resistance is comparable to the intrinsic thermal boundary resistance when phonons are in local thermal equilibrium, and should be taken into consideration in the designs of 2D material devices. Our work also provides important guidelines for correct interpretation and analysis of Raman measurements of heat transport across 2D material interfaces.

# METHODS

**Sample Preparation**

Graphene used in the studies, purchased from Graphene Supermarket, was grown by CVD on copper foil with the back-side etched. Graphene transfer onto the oxide was performed at National University of Singapore following procedures stated in Ref. 66 and Ref. 24. Deposition of single-layer $MoS_2$ onto $SiO_2$/Si substrates using a CVD process[67, 68] and transfer of single-layer $MoS_2$ onto oxide films were performed at Stanford University following the procedures described in Ref. 69. Both top and bottom oxide films are 12-16 nm thick. The $HfO_2$ and $Al_2O_3$ films in our samples were deposited by ALD at a rate of $\approx$1.3 Å/cycle by alternating trimethylaluminum (TMA) and $H_2O$ pulses at 150 °C. The bottom $SiO_2$ oxide layer was deposited by rf magnetron sputtering at a deposition rate of $\approx$0.25 Å/s with a base pressure of $\approx 2\times10^{-7}$ Torr, whereas the top $SiO_2$ oxide layer was deposited by e-beam evaporation at a rate of $\approx$0.2 Å/s with a base pressure of $\approx 10^{-7}$ Torr. The transducer of Al film was deposited by rf magnetron sputtering with a deposition rate of $\approx$0.67 Å/s with a base pressure of $\approx 2\times10^{-7}$ Torr.

**Thickness Determination**

To determine the thickness of oxide films, an additional silicon substrate was included during the deposition of the oxide films. Before being transferred into the chambers, the silicon substrate was soaked in diluted HF for 30 s to remove the native oxide on silicon. After the oxide deposition, thickness of the oxide films is determined by ellipsometry. To determine the thickness of Al transducer layer, a sapphire substrate was included during the deposition of the Al film. The thickness of Al film on sapphire is determined by picosecond acoustics.[70]

**Differential TDTR measurements**

We measured the thermal resistance of oxide/MoS$_2$/oxide and oxide/Gr/oxide interfaces by differential TDTR (see Supporting Information section S2). We employed a modulation frequency of 9.8 MHz, a 5× objective lens with 1/e$^2$ radii of ≈8.4 μm and a total laser power of ≈50 mW, creating a steady-state temperature rise of < 10 K in each measurement. We conducted two conventional TDTR measurements[71] on the regions with and without MoS$_2$ (or graphene) of our samples. We extract $R_{OMO}$ and $R_{OGO}$ by comparing measurements to calculations of a thermal model,[72] see section S2.1 of the Supporting Information for details of our analysis. We note that in our analysis, the $R_{OMO}$ or $R_{OGO}$ is the only fitting parameter, and all other parameters are obtained either from the literature or from stand-alone measurements. Our differential TDTR measurement is mainly sensitive to $R$ of MoS$_2$ and graphene interfaces, and not as sensitive to other input parameters, see Figs. S3(b), ensuring that we achieve accurate measurements of $R_{OMO}$ or $R_{OGO}$ with an uncertainty of ≈8%.

## SUPPORTING INFORMATION

Transfer of single-layer MoS$_2$ and graphene onto oxide films. Differential TDTR approach. Two-dimensional Debye model for $C_{LT}$ calculations for single-layer MoS$_2$. Calculations of effective relaxation time for non-equilibrium LA and TA phonons. Phonon radiation limit for thermal boundary resistance ($R_{rad}$).

## ACKNOWLEDGEMENTS

The work at NUS was supported by the Singapore Ministry of Education Academic Research Fund Tier 2, under Award No. MOE2019-T2-2-135, and Singapore Ministry of Education Academic Research Fund Tier 1 FRC Project FY2016. Sample characterization was carried out in part in the Centre for Advanced 2D Materials. The work at Stanford was supported by in part by ASCENT, one of six centres in JUMP, a Semiconductor Research

Corporation (SRC) program sponsored by DARPA. C.J.M. also acknowledges support from the NSF Graduate Research Fellowship.

# Supporting Information

# Non-equilibrium Phonon Thermal Resistance at MoS$_2$/Oxide and Graphene/Oxide Interfaces

Weidong Zheng[1,2], Connor J. McClellan[3], Eric Pop[3,4], Yee Kan Koh[1,2*]

[1]*Department of Mechanical Engineering, National University of Singapore, Singapore 117576*

[2]*Centre for Advanced 2D Materials, National University of Singapore, Singapore 117542*

[3]*Department of Electrical Engineering, Stanford University, Stanford, California 94305, USA*

[4]*Department of Materials Science and Engineering, Stanford University, Stanford, California 94305, USA*

*Corresponding Author. Email: mpekyk@nus.edu.sg

# S1: Transfer of single-layer MoS$_2$ and graphene onto oxide films

Single-layer MoS$_2$ used for transfer in the preparation of our samples was grown directly onto SiO$_2$/Si substrates by chemical vapor deposition (CVD).[1] For MoS$_2$ transfer, we first spin-coat polypropylene carbonate (PPC) and then polydimethylsiloxane (PDMS) on the MoS$_2$/SiO$_2$/Si. We then soak the stack in de-ionized (DI) water for 30 min. After the PDMS/PPC/MoS$_2$ film is peeled off from the SiO$_2$/Si substrate, we transfer this PDMS/PPC/MoS$_2$ to the substrates of interest to achieve the samples shown in main text Fig. 1(a). Finally, we peel off the PDMS/PPC from MoS$_2$/substrate and achieve clean transfer of single-layer MoS$_2$, see Fig. S1(a).

Graphene in our samples was purchased from Graphene Supermarket, and was grown on copper foil by CVD. For graphene transfer, we follow the procedures described in Refs. 2 and 3. Briefly, we first spin PC (1 wt. % PC in chloroform solution) onto graphene/copper as the support layer. Subsequently, we etch the copper substrate with 7 wt. % ammonium persulfate solution. We clean off the residual etchant on PC/graphene films by placing them on DI water for 5 hour and then transfer them to our oxide substrates. After the PC/graphene/oxide samples are baked on a hot plate at 45°C for 4 min, we dissolve the PC scaffold in chloroform for 24 hours. Finally, we rinse the samples in isopropanol alcohol and blow to dry them using nitrogen. We verify that our graphene transfer is clean, see Fig. S1(b).

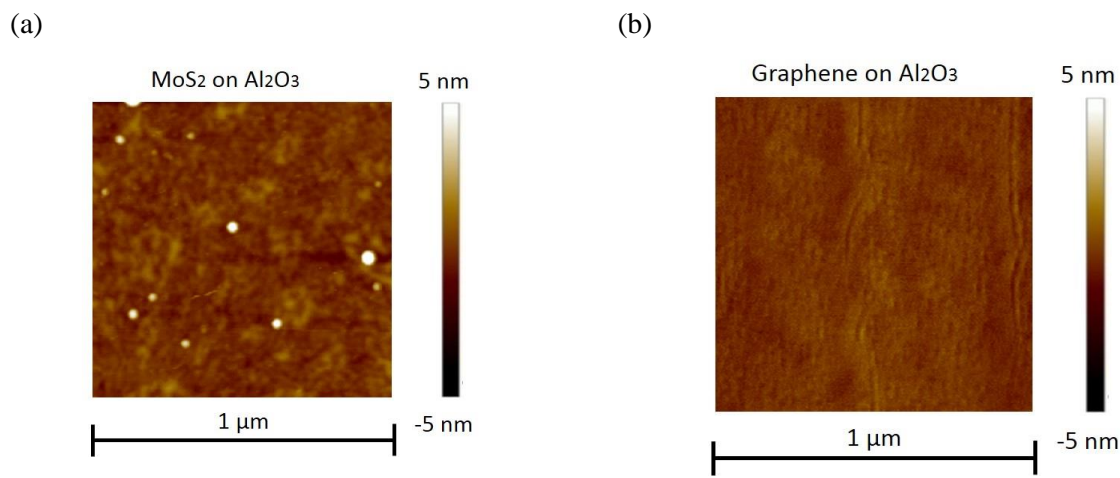

**Fig. S1.** Representative topographic images of the transferred **(a)** MoS$_2$ and **(b)** graphene on Al$_2$O$_3$/GaN, acquired by tapping mode atomic force microscopy. No significant amount of residue is observed in the topographic images, indicating the cleanness of interfaces of our transferred MoS$_2$ and graphene.

## S2: Differential TDTR approach

### S2.1 Differential TDTR measurements of *R* of graphene or MoS₂ interfaces

For differential TDTR measurements, we perform two conventional TDTR measurements on the region with (primary region) and without (control region) the single-layer MoS$_2$ or graphene, see main text Fig. 1(a). From the first measurement on the control region, we derive the thermal conductivity ($\Lambda_{oxide}$) of oxide layers by comparing the ratio ($r_{w/o}$, black symbols in Fig. 2) of in-phase $V_{in}$ and out-of-phase $V_{out}$ signals of the radio-frequency (rf) lock-in amplifier to calculations of a thermal model.[4] In this analysis, $\Lambda_{oxide}$ is the only fitting parameter, and all the other input parameters are determined from the literature or independent measurements, see the fitting (black solid lines) in Fig. S2 The derived thermal conductivity for SiO$_2$, HfO$_2$, Al$_2$O$_3$ are 1.5 W m$^{-1}$ K$^{-1}$, 0.9 W m$^{-1}$ K$^{-1}$, 1.8 W m$^{-1}$ K$^{-1}$, respectively. From the second measurement, we determine the ratio signal ($r_w$, red symbols in Fig. 2) of our TDTR measurement on the primary region. We extract the thermal resistance ($R$) of oxide/MoS$_2$/oxide or oxide/graphene/oxide interfaces by fitting the signal directly using the $\Lambda$ of oxide layers previously derived, see red solid lines in Fig. S2. In the fitting, $R$ of the graphene or MoS$_2$ interfaces is the only free parameter.

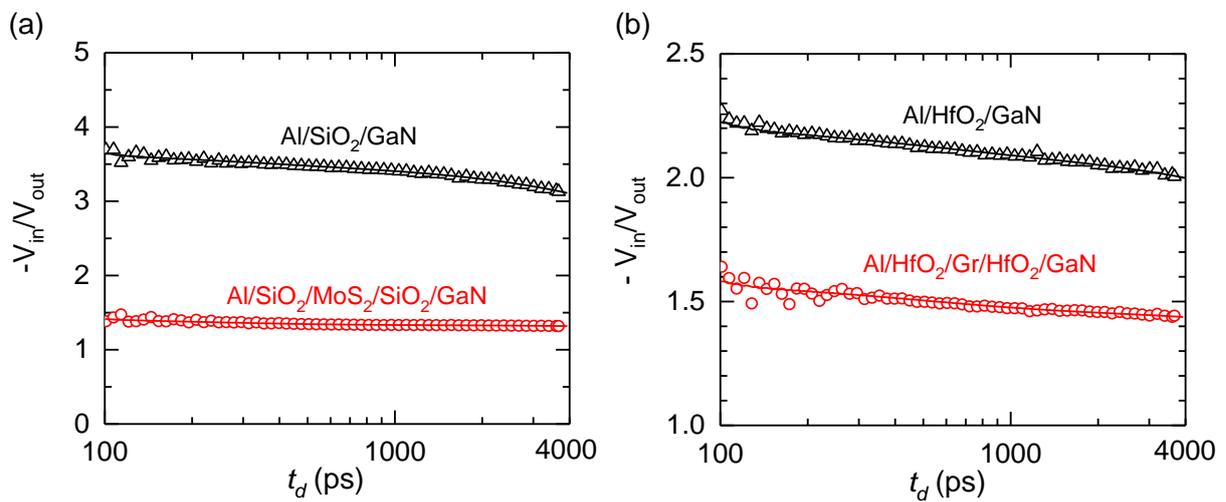

**Fig. S2.** Ratios of in-phase $V_{in}$ and out-of-phase $V_{out}$ TDTR signals as a function of delay time $t_d$ for measurements on the regions with (red symbols) and without (black symbols) **(a)** MoS$_2$ and **(b)** graphene. The solid lines are calculations of a thermal model[4] which is used to fit the measurements.

**S2.2 Uncertainty analysis of differential TDTR measurements**

In our differential TDTR approach, we perform two TDTR measurements, i.e., the first measurement on the control region (without $MoS_2$ or graphene) and the second on the primary region (with $MoS_2$ or graphene). In the first measurement, all the uncertainties from input parameters are lumped into a parameter (e.g., $\Lambda_{oxide}$), by fitting the model calculations to the measured ratio ($r_{w/o}$) of in-phase $V_{in}$ and out-of-phase $V_{out}$ TDTR signals with $\Lambda_{oxide}$ as the only fitting parameter. Note that the value of $\Lambda_{oxide}$ is not meant to be close to the real thermal conductivity of the oxide, but to absorb the errors from all the other input parameters (e.g., thickness of Al film, $R$ of Al/oxide and oxide/GaN interface) in the model. In the second measurement on the primary region (with $MoS_2$ or graphene), we derive the thermal resistance of oxide/$MoS_2$/oxide ($R_{OMO}$) or oxide/Gr/oxide ($R_{OGO}$) interfaces by comparing the measured TDTR ratio signal ($r_w$) to the calculations of the thermal model, using the previously derived $\Lambda_{oxide}$ and the same input parameters as in the fitting of $r_{w/o}$. We note that all input parameters in the fitting of $r_{w/o}$ and $r_w$ should have the same uncertainties, because both regions are on the same substrate going through exactly the same preparation processes. So, as long as the ratio of sensitivities to the parameters are similar in both fitting (which is usually the case as shown in Fig. S3(a)), the use of the derived value of $\Lambda_{oxide}$ in the second fitting of $r_w$ should significantly reduce the uncertainties of $R_{OMO}$ or $R_{OGO}$ derived from the second fitting, see the following discussion.

Here, we quantitatively estimate the uncertainty of our differential TDTR measurements. We first consider the measurement on the primary region, $r_w$, as

$$r_w = f(\alpha_1, \alpha_2, ..., \alpha_N, R) \tag{S1}$$

where $\alpha_i$ are input parameters for both $r_w$ and $r_{w/o}$ (e.g., thermal resistance of Al/oxide interfaces, $\Lambda_{oxide}$ of oxide films in main text Fig. 1(a)), $N$ is the total number of input parameters, and $R$ is the thermal boundary resistance that we intend to derive. We set $\alpha_k$ as the parameter that we lumped all the uncertainty into in the measurement on the control region, i.e., $\Lambda_{\text{oxide}}$. We assume that all the parameters except $\alpha_k$ are independent of the variable $\alpha_i$, and the partial derivative of $r_w$ with respect to $\alpha_i$ is thus

$$\frac{\partial r_w}{\partial \alpha_i} = \frac{\partial f}{\partial \alpha_k}\frac{\partial \alpha_k}{\partial \alpha_i} + \frac{\partial f}{\partial \alpha_i} \tag{S2}$$

We then consider the additional measurement on the control region $r_{w/o}$ as

$$r_{w/o} = g(\alpha_1, \alpha_2, ..., \alpha_N, \gamma) \tag{S3}$$

where $\gamma$ represents the input parameter that is only included in the thermal model for $r_{w/o}$ but not $r_w$, e.g., thermal resistance of oxide/oxide ($R_{\text{oxide/oxide}}$). (Note that we assume $R_{\text{oxide/oxide}}$ is negligible in our analysis.) Mathematically, the additional measurement on the control region can be understood as a constraint condition, i.e., the calculated TDTR ratio signal ($r_{w/o}$) for control region should be treated as a constant, such that the partial derivatives of $r_{w/o}$ with respect to the variable $\alpha_i$ is

$$0 = \frac{\partial g}{\partial \alpha_k}\frac{\partial \alpha_k}{\partial \alpha_i} + \frac{\partial g}{\partial \alpha_i} \tag{S4}$$

Here again we assume that all parameters except $\alpha_k$ are independent of $\alpha_i$. Substituting Eq. (S4) into Eq. (S2), we derive the sensitivity of differential TDTR measurements to $\alpha_i$, $S_{\alpha_i}^d$, as

$$S_{\alpha_i}^d = S_{w,\alpha_i}(1 - \frac{S_{w,\alpha_k}}{S_{w/o,\alpha_k}}\frac{S_{w/o,\alpha_i}}{S_{w,\alpha_i}}) \tag{S5}$$

where $S_{w,\alpha_i}$ and $S_{w,\alpha_k}$ are the sensitivity of $r_w$ signal to $\alpha_i$ and $\alpha_k$, respectively, in the conventional TDTR analysis, $S_{w/o,\alpha_k}$ is the sensitivity of $r_{w/o}$ signal to $\alpha_k$. (The definition of the sensitivity of TDTR ratio signal to the input parameter can be found in Ref. 5.) Note that $S_{\alpha_i}^d$ could be significantly reduced from $S_{w,\alpha_i}$ due to the negative term in Eq. S5 if $\frac{S_{w,\alpha_k}}{S_{w/o,\alpha_k}} \approx \frac{S_{w,\alpha_i}}{S_{w/o,\alpha_i}}$, which is often the case, see Fig. S3(a).

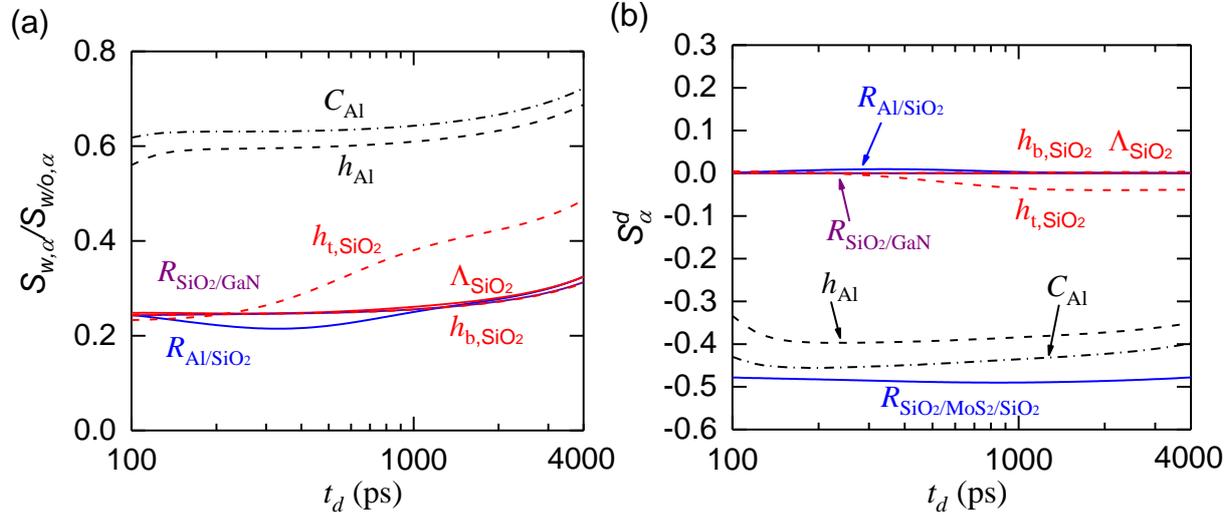

**Fig. S3.** (a) Comparison of $S_{w,\alpha}/S_{w/o,\alpha}$ for various parameters. (b) Sensitivity of differential TDTR signals ($r_w$) to various parameters used in the thermal model for SiO$_2$/MoS$_2$/SiO$_2$ sample. The parameters we include are thermal resistance of SiO$_2$/MoS$_2$/SiO$_2$ ($R_{SiO_2/MoS_2/SiO_2}$), HfO$_2$/Gr/HfO$_2$ ($R_{HfO_2/Gr/HfO_2}$), thermal resistance of oxide/GaN ($R_{oxide/GaN}$) and Al/oxide ($R_{Al/oxide}$) interfaces, thermal conductivity ($\Lambda_{oxide}$) of oxide film, thickness of top ($h_{t,oxide}$) and bottom ($h_{b,oxide}$) oxide layer, thickness ($h_{Al}$) and volumetric heat capacity ($C_{Al}$) of Al film.

Similarly, the sensitivity of TDTR signal $r_w$ to the thermal boundary resistance $R$ that we intend to derive can be expressed as

$$S_R^d = S_{w,R} \tag{S6}$$

where $S_{w,R}$ is the sensitivity of TDTR measurements on the primary region, $r_w$, to $R$. Moreover, in differential TDTR approach, the sensitivity of $r_w$ to $\gamma$ can be determined as

$$S_\gamma^d = -\frac{S_{w,\alpha_k}}{S_{w/o,\alpha_k}} S_{w/o,\gamma} \tag{S7}$$

where $S_{w/o,\gamma}$ is the sensitivity of $r_{w/o}$ signal to $\gamma$ in the conventional TDTR analysis.

Another source of the uncertainty is the determination of phase $\phi_w$ and $\phi_{w/o}$ in the reference channel of the rf lock-in amplifier when we perform measurements on the primary and control region. According to the Ref. 5, in conventional TDTR measurement the sensitivity of TDTR ratio signal $r_w$ to the phase $\phi_w$ is estimated to be $S_{w,\phi} = \dfrac{1}{r_w} \dfrac{\partial r_w}{\partial \phi}$. Similar to the analysis for parameter $\alpha_i$ in differential TDTR measurement, the sensitivity of the TDTR ratio signal to phase $\phi_w$ is

$$S^d_{\phi_w} = S_{w,\phi} \tag{S8}$$

The sensitivity of the TDTR ratio signal to the phase $\phi_{w/o}$ is expressed as

$$S^d_{\phi_{w/o}} = S_{w/o,\phi} \dfrac{S_{w,\alpha_k}}{S_{w/o,\alpha_k}} \tag{S9}$$

where $S_{w/o,\phi}$ is the sensitivity of $r_{w/o}$ signal to $\phi_{w/o}$ in the conventional TDTR analysis.

Thus, the uncertainty of the measurement of $R$ by our differential TDTR can be routinely determined by

$$\dfrac{\Delta R}{R} = \sqrt{\sum_{i=1}^{N} \left(\dfrac{S^d_{\alpha_i}}{S^d_R} \cdot \dfrac{\Delta \alpha_i}{\alpha_i}\right)^2 + \left(\dfrac{S^d_\gamma}{S^d_R} \cdot \dfrac{\Delta \gamma}{\gamma}\right)^2 + \left(\dfrac{S^d_{\phi_{w/o}}}{S^d_R} \cdot \Delta \phi_{w/o}\right)^2 + \left(\dfrac{S^d_{\phi_w}}{S^d_R} \cdot \Delta \phi_w\right)^2} \tag{S10}$$

Together with the sensitivity of differential TDTR ratio signal to each parameter and the uncertainties of input parameters ($\Delta \alpha/\alpha$, $\Delta \gamma/\gamma$, $\Delta \phi_w$ and $\Delta \phi_{w/o}$), we can estimate the uncertainty of the measured $R$ by our differential TDTR approach.

Using the differential TDTR approach, we significantly reduce the sensitivity of our measurements to all parameters, see Fig. S3(b). If we assume uncertainties of ≈5% and ≈3% for $h_{Al}$ and $C_{Al}$, the uncertainty of our measured $R$ calculated using Eq. 10 is ≈8%, significantly lower than if a single TDTR measurement is employed.

## S2.3 Experimental demonstration of the differential TDTR approach

To validate our differential TDTR approach, we fabricate a multilayer structure with a 150-nm-thick $Ni_{80}Fe_{20}$ alloy film sandwiched between two layers of ≈ 40-nm-thick Pd films on a sapphire substrate, with ≈ 80-nm-thick Al as the transducer film. We choose $Ni_{80}Fe_{20}$ alloy film for the validation because it has been well studied in our previous work[6] and we can independently measure the thermal conductivity of this film from the in-plane electrical resistivity using the Wiedemann-Franz law. All of these thin films are deposited through thermal evaporation under a base pressure of ≈ $10^{-7}$ Torr. We determine the thickness of these metal films by picosecond acoustics.[7]

We first perform one TDTR measurement on the control region (the region without $Ni_{80}Fe_{20}$ alloy film). With thermal conductance $G_{Al/Pd}$ = 105 MW m$^{-2}$ K$^{-1}$ of Al/Pd interfaces as input parameter and $G_{Pd/sapp}$ of Pd/sapphire interfaces as the only free parameter, we derive $G_{Pd/sapp}$ = 132 MW m$^{-2}$ K$^{-1}$ by comparing our measured TDTR ratio signal to the calculations of the thermal model, see the black symbols and solid line in Fig. S4. We then perform another TDTR measurement on the primary region (the region with $Ni_{80}Fe_{20}$ alloy film). Together with $G_{Al/Pd}$ and $G_{Pd/sapp}$ determined previously, we fit $r_w$ with thermal conductivity ($\Lambda_a$) of $Ni_{80}Fe_{20}$ alloy film as the only free parameter, as shown by red symbols and solid line in Fig. S4, and derive $\Lambda_a$ = 20.3±3.0 W m$^{-1}$ K$^{-1}$, in good agreement with the measurements in our previous work[6] and that by four-point probe measurement, see Table 1 and the following discussions. We note that in this measurement we independently measure thermal conductance of Pd/$Ni_{80}Fe_{20}$ interfaces to be 700 MW m$^{-2}$ K$^{-1}$ with an uncertainty of ≈40%.

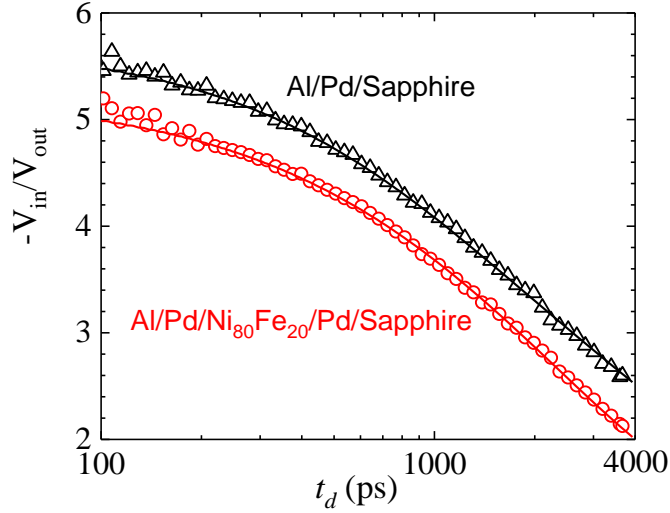

**Fig. S4:** Fitting of the ratio of in-phase $V_{in}$ and out-of-phase $V_{out}$ TDTR signals measured on the control region (black symbols) and primary region (red symbols) of $Ni_{80}Fe_{20}$ sample as a function of delay time $t_d$. The solid lines are the calculations of the thermal model which is used to fit the measurements.

To verify the measured $\Lambda_a$ of $Ni_{80}Fe_{20}$ film by our differential TDTR, we also derive $\Lambda_a$ of $Ni_{80}Fe_{20}$ alloy from electrical resistivity $\rho$ using Wiedeman-Franz law $\Lambda = LT/\rho$, where $L$ is the Lorenz number for metal film, $T$ is the temperature. We determine the electrical resistivity $\rho$ = 37.1±2.2 μΩcm of our $Ni_{80}Fe_{20}$ film on the glass slide by a four-point probe. To include the possible 2~3 W m$^{-1}$ K$^{-1}$ of heat carried by phonons in alloys,[8] we follow the procedure stated in our previous study[6] and cite the Lorenz number specifically for $Ni_{80}Fe_{20}$ alloy as (2.38±0.2)×10$^{-8}$ Ω W K$^{-2}$. We estimate the thermal conductivity of our $Ni_{80}Fe_{20}$ film to be 19.2±2.0 W m$^{-1}$ K$^{-1}$, consistent with the measurements in our previous work.[6] We summarize our measurements by differential TDTR, conventional TDTR and four-point measurements in Table. 1.

**Table 1**. Summary of measured thermal conductivity ($\Lambda_a$) of 150-nm-thick $Ni_{80}Fe_{20}$ film by different approaches.

|  | Differential TDTR | Conventional TDTR | Four-point |
|---|---|---|---|
| $\Lambda_a$ (W m$^{-1}$ K$^{-1}$) | 20.3±3.6 | 27±12.8 | 19.2±2.0 |

## S3: Two-dimensional Debye model for $C_{LT}$ calculations for single-layer MoS$_2$

To determine the heat capacity per unit area $C_{LT}$ (unit: J m$^{-2}$ K$^{-1}$) of LA and TA phonons in single-layer MoS$_2$, we build a two-dimensional (2D) Debye model similar to the models described in Refs. 9-11. We assume a linear dispersion for the LA and TA branches; $\omega = v_{L(T)} \times q$, where $\omega$ is the phonon frequency, $v_L = 6850$ m s$^{-1}$ and $v_T = 5372$ m s$^{-1}$ are the speeds of sound[12] of LA and TA modes, respectively, and $q$ is the phonon wavevector. For the ZA branch, we assume a quadratic dispersion; $\omega = \alpha \times q^2$, where $\alpha = 5.23 \times 10^{-7}$ m$^2$ s$^{-1}$ is a parameter fitted from the phonon dispersion. With the assumptions, the 2D phonon density of states (PDOS) for each branch $j$, $D_j(\omega)$, are derived as $D_j(\omega) = \frac{\omega}{2\pi v_j^2}$ for LA and TA branches, and $D_j(\omega) = \frac{1}{4\pi\alpha}$ for the ZA branch.[9-11] Thus, $C_{LT}$ can be evaluated from

$$C_{LT} = \sum_j \int_0^{\omega_{m,j}} D_j(\omega) \hbar\omega \frac{\partial f}{\partial T} d\omega \tag{S1}$$

where $j$ includes LA and TA branches, $f$ is Bose-Einstein distribution, $\hbar$ is the reduced Plank constant, $T$ is the temperature, $\omega_{m,j}$ refers to maximum phonon frequency of phonon branch $j$. To ensure the correct total number of acoustic modes, $\omega_{m,j}$ is derived from the planar primitive cell density ($n$) using $\int_0^{\omega_{m,j}} D_j(\omega) d\omega = n$. Thus, $\omega_{m,j} = \sqrt{4\pi n} v_j$ for LA, TA branch and $\omega_{m,Z} = 4\pi\alpha n$ for ZA branch. We approximate the heat capacity of the optical phonons from the values in Ref. 13. We first determine the ratio of heat capacity from optical phonons to the total heat capacity, using the equation $C_t = \int D_A(\omega) \hbar\omega \frac{\partial f}{\partial T} d\omega + \int D_O(\omega) \hbar\omega \frac{\partial f}{\partial T} d\omega$, where $D_A$ and $D_O$ are the PDOS for acoustic and optical phonons, respectively, determined from Fig. 4(a) in Ref. 13. Together with the calculated total heat capacity from Fig. 2(a) in Ref. 13, we derive the heat capacity of the optical phonons.

We verify the accuracy of our 2D Debye model by comparing the total heat capacity ($C_t$) of MoS$_2$ to that determined by molecular dynamics (MD) simulations[13] and first principles calculations,[14] see Fig. S5. We find that calculations of our 2D Debye model deviate from the calculations in Refs. 13 and 14 by < 10% for high temperature (> 200 K) and by 10%-20% below 200 K. Despite the deviation, conclusions reached in this work are still reliable because we mainly discuss the measurements over the temperature range 300-600 K in Figs. 2 (a) and 3(a).

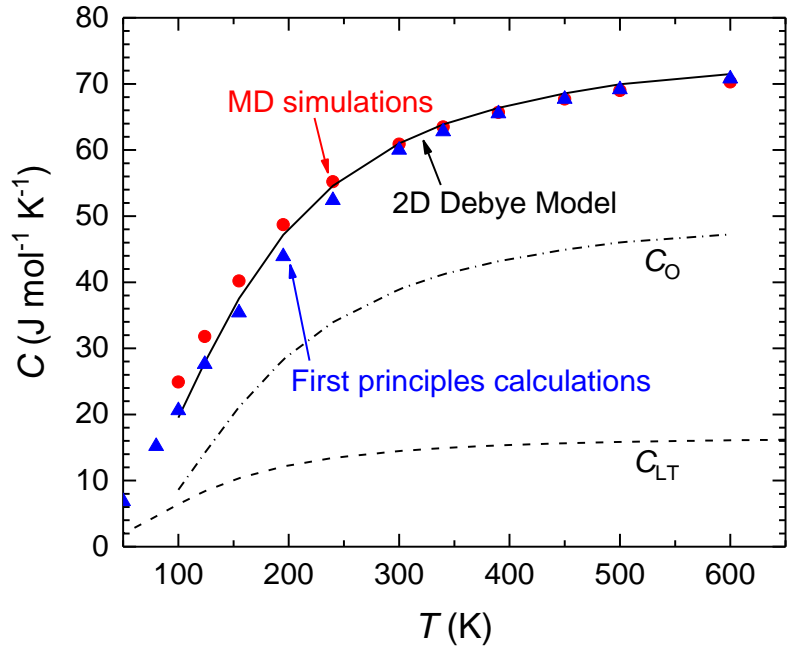

**Fig. S5.** Comparison of total heat capacity of $MoS_2$ determined by MD simulations (red circles), first principles calculations (blue up triangles) and 2D Debye model (solid line). The dashed line represents the heat capacity of LA and TA phonons calculated by 2D Debye model. The dashed dot line represents the heat capacity of the optical phonons approximated from the total heat capacity and PDOS of acoustic and optical phonons reported in Ref. 13.

# S4: Calculations of effective relaxation time for non-equilibrium LA and TA phonons

To estimate the effective relaxation time ($\tau_{\text{eff}}$) for the anharmonic scattering process between the in-plane vibration modes (i.e., LA and TA phonons) and ZA modes, we first group the LA and TA phonons into multiple channels, and all phonons in each channel $i$ are assumed to have the same frequency $\omega_i$ and relaxation time $\tau(\omega_i)$, with heat capacity of $C(\omega_i) = D(\omega_i)\hbar\omega_i \frac{\partial f}{\partial T} \Delta\omega_i$, where $D(\omega_i)$ denotes the density of states for phonons in channel $i$, $\Delta\omega_i$ is the frequency interval of channel $i$, and $f$ is Bose-Einstein distribution. We assume that energy is transferred independently and in parallel from phonons in each channel $i$ directly to ZA phonons. In other words, the total heat transport from LA and TA modes to ZA modes is the sum of contribution from each channel, $\frac{D(\omega_i)}{\tau(\omega_i)}\hbar\omega_i \frac{\partial f}{\partial T} \Delta\omega_i$. With the assumption, $\tau_{\text{eff}}$ is the harmonic average of the relaxation time $\tau(\omega_i)$, weighted by the heat capacity $C(\omega_i)$ of each channel, i.e.,

$$\frac{C_{LT}}{\tau_{\text{eff}}} = \sum_i \frac{D(\omega_i)}{\tau(\omega_i)} \hbar\omega_i \frac{\partial f}{\partial T} \Delta\omega_i \tag{S2}$$

where, $C_{LT}$ is the heat capacity of all LA and TA phonons, as in the main text.

In our calculations, we obtain $\tau(\omega_i)$ from calculations in the literature. Thus, the frequency intervals of our channels are not uniform, depending on the $\tau(\omega_i)$ data available in the literature. We thus use $\Delta\omega_i = (\omega_i - \omega_{i-1})/2 + (\omega_{i+1} - \omega_i)/2$ to calculate $\Delta\omega_i$. For single-layer graphene, we approximate $\tau(\omega_i)$ from the relaxation time of LA and TA modes in supported graphene on $SiO_2$ extracted from molecular dynamics simulations and spectral energy density analysis in Ref.15. We determine $D(\omega_i)$ from the calculations by a lattice dynamics model reported in Ref. 9. For single-layer $MoS_2$, we estimate $\tau(\omega_i)$ from the relaxation time of LA and TA modes in bulk $MoS_2$ determined by first principles calculations in Ref. 16, while we calculate $D(\omega_i)$ from the 2D Debye model as introduced in Section S3. (We note that previous experiments and calculations suggest that the relaxation times of phonons in single-layer 2D materials and in bulk structures are on the same order of magnitude.[17, 18] Our simple model is only meant to give a crude "order-of-magnitude" approximation of $R_{int}$, and thus an accurate value of $\tau_{eff}$ is not vital.) We thus determine $\tau_{eff} \approx 5.0$ ps and 8.8 ps for graphene and MoS$_2$, respectively, at $T = 300$ K.

We treat the temperature dependence of the relaxation time as a fitting parameter and find that if we assume $\tau^{-1}(\omega_i) \propto T$,[19-23] our calculations of $R_{ne}$ fit the Raman measurements well. Thus, $\tau_{eff}$ would exhibit a similar behavior, $\tau_{eff}^{-1} \propto T$, which is used in the calculations of $R_{ne}$ in Fig. 3.

## S5: Phonon radiation limit for thermal boundary resistance ($R_{rad}$)

In the literature, phonon radiation limit for the thermal boundary conductance ($G_{rad}$) is well documented. Here, we derive $R_{rad}$ in the main text from $G_{rad}$ using $R_{rad} = 1/G_{rad}$. We calculate $G_{rad}$ of MoS$_2$/substrate and Gr/substrate interfaces following the Eqs. (2) and (3) reported in Ref. 24. In the calculations, we approximate the phonon flux ($h_{2D}$) in MoS$_2$ and graphene using our anisotropic model with a truncated linear dispersion of MoS$_2$ and graphite.[12] The input parameters used in our calculations of $h_{2D}$ are listed in Table 3 in Ref. 12. For phonon flux in substrates, we employ the same equation as described in Ref. 24, while we approximate $\omega_{m,j} = k_B \Theta_{eff}/\hbar$ for LA and TA phonon branches in substrates.

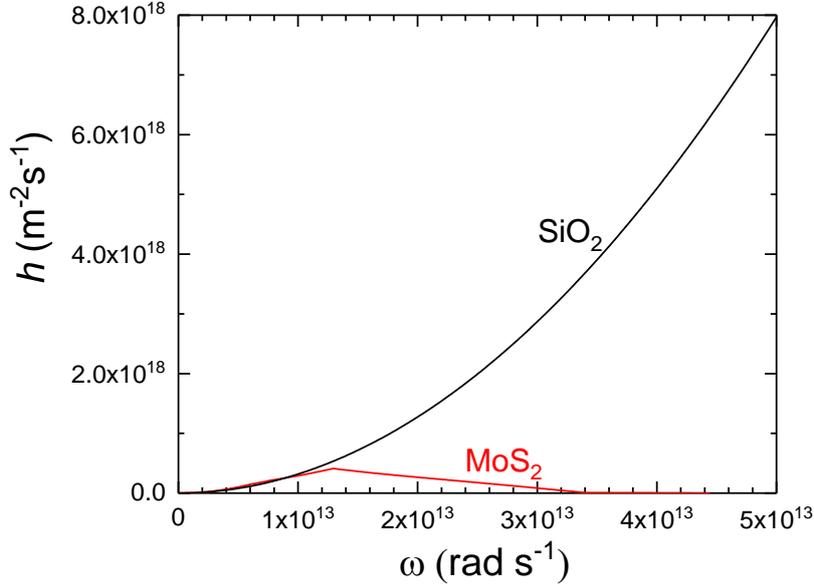

**Fig. S6**. Comparison of phonon flux of MoS$_2$ and SiO$_2$

Temperature affects heat transport across interfaces by elastic phonon transmission in three aspects: phonon heat capacity (phonon population and phonon energy distribution), group velocity and transmission probability.[25] Since $h_{SiO2} \gg h_{MoS2}$ for a large portion of phonons (see Fig. S6), the transmission probability of most phonons leaving MoS$_2$ is thus approximate 100%.[25] As a result, thermal boundary resistance ($R$) of MoS$_2$/SiO$_2$ interfaces is dominated by the product of heat capacity and group velocity of acoustic phonons in MoS$_2$ (optical phonons play a negligible role in heat transport due to the fairly low group velocity[12]). In general, the phonon group velocity is independent of temperature[25] and thus the derived temperature dependence of $R$ for $MoS_2/SiO_2$ interfaces primarily originates from the temperature dependence of heat capacity of acoustic phonons of $MoS_2$. Since the acoustic Debye temperature of $MoS_2$ is only ≈260 K,[26] the acoustic phonons in $MoS_2$ are fully excited and heat capacity shows a fairly weak temperature dependence when $T > 300$ K. Therefore, a weak temperature dependence of $R$ is expected over the temperature range >300 K in Fig. 3(a).